\documentstyle[12pt,aaspp4]{article}

\def\la{\mathrel{\hbox{\rlap{\hbox{\lower4pt\hbox{$\sim$}}}\hbox{$<$}}}}
\def\ga{\mathrel{\hbox{\rlap{\hbox{\lower4pt\hbox{$\sim$}}}\hbox{$>$}}}}

\slugcomment{Accepted to {\it The Astrophysical Journal}}

\received{2005 May 23}
\begin{document}

\title{Early-Time Chromatic Variations in the Wind-Swept Medium of GRB 021211 and the Faintness of its Afterglow}

\author{M. C. Nysewander\altaffilmark{1}, D. E. Reichart\altaffilmark{1},
H.-S. Park\altaffilmark{2}, G. G. Williams\altaffilmark{3}, 
K. Kinugasa\altaffilmark{4}, D. Q. Lamb\altaffilmark{5},
A. A. Henden\altaffilmark{6}, S. Klose\altaffilmark{7}, 
T. Kato\altaffilmark{8}, A. Harper\altaffilmark{5}, H. Yamaoka\altaffilmark{9},
C. Laws\altaffilmark{10}, K. Torii\altaffilmark{11}, D. G.
York\altaffilmark{5},
J. C. Barentine\altaffilmark{12}, J. Dembicky\altaffilmark{12}, 
R. J. McMillan\altaffilmark{12}, J. A. Moran\altaffilmark{1},
D. H. Hartmann\altaffilmark{13}, B. Ketzeback\altaffilmark{12}, 
M. B. Bayliss\altaffilmark{1}, J. W. Bartelme\altaffilmark{1},
J. A. Crain\altaffilmark{1}, A. C. Foster\altaffilmark{1}, 
M. Schwartz\altaffilmark{14}, P. Holvorcem\altaffilmark{14}, 
P. A. Price\altaffilmark{15}, R. Canterna\altaffilmark{16},
G. B. Crew\altaffilmark{17}, G. R. Ricker\altaffilmark{17}, and 
S. D. Barthelmy\altaffilmark{18}
}

\altaffiltext{1}{Department of Physics and Astronomy, University of North
Carolina at Chapel Hill, Campus Box 3255, Chapel Hill, NC 27599;
mnysewan@physics.unc.edu, reichart@physics.unc.edu}
\altaffiltext{2}{Lawerence Livermore National Laboratory, 7000 East Avenue,
Livermore, CA 94550}
\altaffiltext{3}{MMT Observatory, University of Arizona, Tucson, AZ 85721}
\altaffiltext{4}{Gunma Astronomical Observatory, 6860-86 Nakayama
Takayama, Agatsuma, Gunma 377-0702, Japan}
\altaffiltext{5}{Department of Astronomy \& Astrophysics, University of
Chicago, 5640 S. Ellis Ave., Chicago, IL, 60615}
\altaffiltext{6}{AAVSO, Clinton B. Ford Astronomical Data and Research Center,
25 Birch St., Cambridge, MA}
\altaffiltext{7}{Th\"{u}ringer Landessternwarte, 07778 Tautenburg, Germany}
\altaffiltext{8}{Department of Astronomy, Faculty of Science, Kyoto University,
Sakyo-ku, Kyoto 606-8502, Japan}
\altaffiltext{9}{Department of Physics, Kyushu University, Fukuoka 810-8560,
Japan}
\altaffiltext{10}{Department of Astronomy, Box 351580, University of 
Washington, Seattle, WA 98195}
\altaffiltext{11}{Department of Earth and Space Science, Graduate School of
Science, 1-1 Machikaneyama-cho, Toyonaka, Osaka 560-0043, Japan}
\altaffiltext{12}{Apache Point Observatory, P.O. Box 59, Sunspot, NM 88349}
\altaffiltext{13}{Clemson University, Department of Physics and Astronomy,
Clemson, SC 29634}
\altaffiltext{14}{Tenagra Observatory, HC2 Box 292, Nogales, AZ 85621}
\altaffiltext{15}{University of Hawaii, Institute of Astronomy, 2680 Woodlawn
Drive, Honolulu, HI 96822-1897}
\altaffiltext{16}{University of Wyoming, Department of Physics and Astronomy,
P.O. Box 3905, Laramie, WY 82072}
\altaffiltext{17}{Center for Space Research, Massachusetts Institute of
Technology, 70 Vassar Street, Cambridge, MA, 02139}
\altaffiltext{18}{NASA Goddard Space Flight Center, Code 661, Greenbelt, MD
20771}

\begin{abstract}

We present Follow-Up Network for Gamma-Ray Bursts (FUN GRB) Collaboration
observations of the optical afterglow of GRB 021211 made between 143 seconds
and 102 days after the burst.  Our unique data set includes the earliest
filtered detections and color information for an afterglow in the pre-Swift
era.  We find that the afterglow is best described by (1) propagation through
a wind-swept medium, (2) a cooling break that is blueward of the observed
optical frequencies, and (3) a hard electron energy distribution.  However,
superimposed on this ``standard model'' behavior we find one and possibly two significant chromatic
variations during the first few hours after the burst.  We consider possible
reasons for these variations, including the possibility that they are due to a dust echo.
Finally, we constrain physical parameters that describe the afterglow and
surrounding medium for a variety of scenarios and find that GRB 021211's
afterglow is faint for a combination of 3 -- 4 reasons:  (1) a low fraction of
energy in relativistic electrons, (2) a low density for the wind-swept medium,
implying either a low mass-loss rate and/or a high wind velocity for the
progenitor, (3) a wide opening/viewing angle for the jet, and possibly (4) moderate source-frame extinction.  The jet appears
to be significantly far from equipartition and magnetically dominated.  More extreme versions of this might explain the darkness of many afterglows in the Swift era.

\end{abstract}


\section{Introduction}

Discovery of GRB afterglows has become almost commonplace.  However, we are still in a regime where nearly every well-sampled afterglow 
contributes to our understanding of the phenomenon in new and meaningful ways.
 Observationally, GRB 021211 distinguishes itself in two ways:  (1) It is the
second GRB for which an optical afterglow was observed within minutes of the
burst, thanks to rapid responses by the HETE-2 satellite (Crew et al. 2002,
2003) and three robotic telescopes -- RAPTOR (Wozniak et al. 2002); KAIT (Li et
al. 2002, 2003); and Super-LOTIS (Park, Williams \& Barthelmy 2002; this
paper); and (2) It is the first GRB for which filtered detections (beginning
143 seconds after the burst) and color information (beginning 38 minutes after
the burst) were obtained at early times.  

In addition to observations presented in GRB Coordinates Network (GCN)
Circulars, many groups have presented their observations in peer-reviewed journals:  Li et al. (2003) present an unfiltered light curve beginning 105
seconds after the burst; Fox et al. (2003) present an unfiltered light curve
beginning 21 minutes after the burst and filtered optical, NIR, and radio
observations beginning 2.0 hours after the burst; Pandey et al. (2003) present
filtered optical observations beginning 6.8 hours after the burst; Holland
et al. (2004) present filtered optical and NIR observations of both the
afterglow and host galaxy beginning 17 hours after the burst and measure the
spectral flux distribution of the afterglow around 21 hours after the
burst; and Smith et al. (2005) present submillimeter observations around 25 hours and 10 days after the burst.  Finally, Della Valle et al. (2003) present photometric and spectral
evidence for an associated supernova at late times.

As in the case of GRB 990123 (Akerlof et al. 1999), the optical afterglow
faded more rapidly at first, presumably due to a reverse shock (Wei 2003; Fox
et al. 2003; Li et al. 2003; Holland et al. 2004).  However, these afterglows
differ in that GRB 021211 was $\approx$3 -- 4 mag fainter, despite a lower
redshift [$z
= 1.004$ for GRB 021211 (Vreeswijk et al. 2002; Della Valle et al. 2002) vs. $z = 1.600$ for GRB 990123
(Hjorth et al. 1999)] (Fox et al. 2003; Li et al. 2003; Pandey et al. 2003;
Crew et al. 2003).  If it were not for the rapid response of the GRB
community, GRB 021211 might have been called a ``dark burst'':  It faded from R
$\approx$ 14 mag at $\approx$90 sec after the burst (Wozniak et al. 2002) to R $>$ 21
mag about
three hours later.  Many bursts that would have been called ``dark'' in the
BeppoSAX era are being and will be called ``dim'' in the HETE-2, Integral, and Swift era
due to faster responses.

Some authors have modeled GRB 021211 with an emphasis on its environment. 
Kumar \& Panaitescu (2003) argue that the GRB and afterglow were produced by
the same shock and within this framework constrain physical parameters for
both constant-density and wind-swept media.  Panaitescu \& Kumar (2004)
consider the early-time afterglows of both GRB 021211 and GRB 990123 in the
context of reverse-forward shock (for both constant-density and wind-swept
media) and wind-bubble scenarios and find that the reverse-forward shock scenario is preferred. 
Chevalier, Li \& Fransson (2004) argue for a wind-swept medium with the cooling
break redward of the R band and within this framework find wind densities that
are low compared to Galactic Wolf-Rayet stars (see also Panaitescu \& Kumar
2004).  Finally, Dado, Dar \& De Rujula (2003) model GRB 021211 within the
framework of their cannonball model.

In \S2, we present FUN GRB Collaboration observations of GRB 021211, which
include the earliest filtered detections and color information for an
afterglow in the pre-Swift era.  In \S3, we fit standard afterglow and
extinction curve models to these and other groups' data and show that within
the first few hours after the burst one and possibly two significant chromatic variations are
superimposed on this ``standard model'' behavior.  In \S4, we compare our
results
to previous modeling results and discuss possible reasons for these chromatic
variations, including the possibility that they are due to a dust echo.
We
also constrain physical parameters that describe the afterglow and surrounding
medium for a variety of scenarios and discuss why GRB 021211's afterglow is
so faint.
We draw conclusions in \S5.

\section{Observations}

Long-duration, X-ray rich GRB 021211 was detected by HETE-2's FREGATE, WXM,
and SXC instruments on December 11, 2002 at 11:18:34 UTC (Crew et al. 2003). 
The initial spacecraft localization was 14$\arcmin$ in radius and reported in
near-real time, only 22 seconds after the burst.  Ground analysis of the WXM
and SXC data, reported 131 minutes after the burst, improved the localization
to 2$\arcmin$ in radius and was consistent with the initial localization.

Fox \& Price (2002) announced the discovery of an R $\sim$ 18 mag and fading,
stationary point source in the error circle 53 minutes after the burst.  While
the pair labored, the robotic telescopes of three groups had already responded
to the alert.  For only the second time in the afterglow era, robotic
telescopes extended the light curve of an afterglow back to within tens of
seconds of the burst (Wozniak et al. 2002; Li et al. 2002, 2003; Park,
Williams \& Barthelmy 2002; this paper).

The dim and quickly fading afterglow soon grew too faint for small telescopes,
and a possible host galaxy was detected (Lamb et al. 2002a, 2002b; McLeod et
al. 2002) but later confirmed under better seeing conditions to be cleanly
separated from the afterglow by 1.5$\arcsec$ (Caldwell et al. 2002).  VLT spectroscopy
of the true host galaxy resulted in a measured redshift of $z = 1.004 \pm 0.002$
(Vreeswijk et al. 2002; Della Valle et al. 2002).  Late-time observations
indicate both a re-brightening at the time expected for a supernova at $z \sim 1$,
and a spectrum that resembles that of Type Ic SN 1994I (Fruchter et al. 2002;
Della Valle et al. 2003).

\subsection{FUN GRB Collaboration Observations}

We summarize FUN GRB Collaboration observations of GRB 021211 in Table 1.  We
have calibrated all of our measurements using the field calibration of Henden
(2002).

Super-LOTIS imaged the entire GRB 021211 field in R band beginning 143 and 309
seconds after the burst (Park, Williams \& Barthelmy 2002).  Super-LOTIS is a
fully automated f/3.5 0.6-meter diameter Perkin-Elmer telescope on a Boller \&
Chivens mount at Kitt Peak National Observatory.  The camera is a 2048 $\times$ 2048
Loral CCD, which yields a large, 51$\arcmin$ $\times$ 51$\arcmin$ field of view.  Observations began
automatically after receiving the HETE-2 alert via a socket connection to the
GCN.  Both exposures were 60 seconds in duration.  The mean times that we list
in Table 1 are flux weighted using an iterated power-law index of $\alpha =
-1.37$, since the exposure time is comparable to the age of the burst, at least for the first exposure. 
This results in small shifts of 2.4 and 1.2 seconds in the mean times of these
observations.

Tenagra Observatories', Ltd., 0.81-meter Tenegra II telescope imaged the GRB
021211 field beginning 37 minutes after the burst.  We obtained four sets of
four images, each set in a 2 $\times$ 2 arrangement to cover the initial 28$\arcmin$-diameter
localization and each in a different filter (I$_{\rm c}$R$_{\rm c}$VB).  We then re-pointed to
the candidate afterglow of Fox \& Price (2002) and cycled through I$_{\rm c}$R$_{\rm c}$ thrice
more.  Of these, we combined the first two I$_{\rm c}$ and R$_{\rm c}$ images to optimize signal
to noise, but the final two images were not usable due to the onset of
morning.  This resulted in three detections (I$_{\rm c}$R$_{\rm c}$V), a limit (B), and two more
detections (I$_{\rm c}$R$_{\rm c}$).  We reduced the images using IRAF's CCDRED package and
performed PSF photometry using IRAF's DAOPHOT package.

We imaged the central 11$\arcmin$ $\times$ 11$\arcmin$ of the initial 28$\arcmin$-diameter localization in R$_{\rm c}$
band beginning 85 minutes after the burst from Gunma Astronomical Observatory,
located in Agatsuma, Gunma, Japan (Kinugasa et al. 2002).  We used the f/12
0.65-meter diameter Cassegrain telescope, which is equipped with an Apogee AP8
1024 $\times$ 1024 back-illuminated SITe CCD.  We obtained a total of 28 images,
which we combined to optimize signal to noise.  We reduced the images using
IRAF's CCDRED package and performed PSF photometry on the combined image using
IRAF's DAOPHOT package.

We reacquired the field with the 1.34-meter diameter Tautenburg Schmidt
telescope 11.7 hours after the burst and imaged in R and I bands for the next
1.1 hours using the 2048 $\times$ 2048 prime-focus CCD (Klose et al. 2002).  However,
we did not detect the afterglow.

We began observations with the 3.5-meter diameter Astrophysics Research
Consortium (ARC) telescope at Apache Point Observatory 22.0 hours after the
burst, and returned to the field on December 28 and March 23, 17 and 102 days
after the burst (Lamb et al. 2002a, 2002b).  All images were taken in i$^*$ band
using SPIcam, a 2048 $\times$ 2048 back-illuminated SITe CCD.  Three 2000-second
images were taken on the first night, and seven 1200-second images were taken
on each of the following nights.  We reduced, combined, and calibrated these
images using IRAF's CCDRED and DAOPHOT packages.

Finally, we re-observed the field on December 13 with the 1.0-meter diameter
telescope at the U.S. Naval Observatory's Flagstaff Station for purposes of
calibration (Henden 2002).  BVR$_{\rm c}$I$_{\rm c}$ images were taken with a 2048 $\times$ 2048
back-illuminated SITe/Tektronix CCD under 2.2$\arcsec$ seeing conditions.  Upon
inspection of the images, the afterglow was still marginally visible in the
8-minute V-band image.  The afterglow was measured using a two-FWHM diameter
aperture with IRAF's DAOPHOT package.

\subsection{Implications of Late-Time ARC Observations}

Supernova signatures had been found for many GRBs prior to GRB 021211 (e.g.,
Galama et al. 1998; Bloom et al. 1999; Reichart 1999; Galama et al. 2000;
Bloom et al. 2002; Garnavich et al. 2003; Price et al. 2003; Stanek et al.
2003; see also Zeh, Klose \& Hartmann 2004 for a systematic analysis).  For GRB 021211, Fruchter et al. (2002) and Della Valle et al. (2003)
found evidence for excess red light $\approx$25 days after the burst, and Della
Valle
et al. (2003) obtained a VLT spectrum at 27 days.  This spectrum exhibits Ca
II absorption with a relative velocity of $\approx$14,440 km/s for $z = 1.004$ and
is
similar to other Type Ic spectra.

Our late-time ARC observations neither confirm nor contradict the existence of this
underlying supernova.  Subtraction of our second and third i$^*$ epochs using
ISIS-2 (Alard 2000) does not reveal any residual flux.  However, this is
likely due to the timing of our observations:  The re-brightening reported by
Della Valle et al. (2003) occurs mostly between our observations at 17 and 102
days after the burst.  In Figure 1, we plot our i$^*$ light curve and the fitted
afterglow model of \S3.1.

\subsection{Recalibration of KAIT and NEAT Photometry}

To better investigate possible chromatic variations that occurred during the
unfiltered KAIT and NEAT observations (see \S3.2), we have recalibrated these
measurements from the R$_{\rm c}$ band to broad bands given by the spectral responses
of their respective CCDs (Pravdo et al. 1999; Li et al. 2003):  W. Li (private
communication) and P. Price (private communication) kindly provided us with
their calibration stars.  Using the BVR$_{\rm c}$I$_{\rm c}$ magnitudes of these stars from
Henden (2002), we fitted extinguished blackbody functions to each of these
stars and then integrated these fitted functions against the appropriate
spectral response curve.  This resulted in small, 0.05 and 0.03 magnitude
offsets in the calibration of the KAIT and NEAT measurements, respectively.

\section{Analysis}

We now fit standard afterglow and extinction curve models to these and other
groups' data and show that within the first few hours after the burst
one and possibly two significant chromatic variations are superimposed on this ``standard model''
behavior.  The data that we include in this analysis are plotted in Figure 2
and consist of FUN GRB Collaboration data (\S2.1), data previously published in
peer-reviewed journals (Pandey et al. 2003; Fox et al. 2003; Li et al. 2003;
Holland et al. 2004), and data from the GCN archive (McLeod et al. 2002). 
These data span the first $\approx$2.5 days after the burst, after which the host
galaxy and supernova become contaminants.  All magnitudes have been converted
to spectral fluxes as prescribed by Bessell (1979) and Bessell \& Brett (1998).

\subsection{Model and Fits}

We now model these data and constrain model parameters.  We model the
afterglow with two components, corresponding to reverse and forward shocks. 
Each component has a power-law light curve and a power-law spectrum, but the
spectrum is extinguished by dust in the source frame and in our Galaxy and
absorbed by hydrogen in the source frame and the Ly$\alpha$ forest:
\begin{equation}
F_\nu(t) = e^{-\tau_\nu^{MW}}e^{-\tau_{\nu (1+z)}^{Ly\alpha}}e^{-\tau_{\nu (1+z)}^{source}}F_0\left[\left(\frac{t}{t_0}\right)^{\alpha_{rs}}\left(\frac{\nu}{\nu_R}\right)^{\beta_{rs}}+\left(\frac{t}{t_0}\right)^{\alpha_{fs}}\left(\frac{\nu}{\nu_R}\right)^{\beta_{fs}}\right],
\end{equation}
where $\tau_\nu^{MW}$ is the Galactic extinction curve model of Cardelli,
Clayton \&
Mathis (1989), $\tau_{\nu (1+z)}^{Ly\alpha}$ is the Ly$\alpha$ forest
absorption model of Reichart
(2001a), $\tau_{\nu (1+z)}^{source}$ is the source-frame extinction curve and
Lyman limit
absorption model of Reichart (2001a), $\alpha_{rs}$ and $\alpha_{fs}$ are the
temporal indices of
the two components, $\beta_{rs}$ and $\beta_{fs}$ are the spectral indices
of the two
components, $\nu_R$ is the effective frequency of the R band, $t_0$ is the time
when
these two components are of equal brightness at this frequency, and $F_0$ is this
brightness.  Since the extinction and absorption models have features that are
narrower than most photometric bands, we integrate Equation 1 against the
appropriate filter transmissivity curve (or CCD spectral response curve for
the unfiltered measurements; \S2.3) before fitting it to the data. 

We fit this model to the data using Bayesian inference (e.g., Reichart 2001a;
Lee et al. 2001; Galama et al. 2003):  The posterior probability distribution
is equal to the product of the prior probability distribution and the
likelihood function.  The likelihood function is given by:
\begin{equation}
{\cal L} = \prod_{i=1}^N \frac{1}{\sqrt{2\pi(\sigma_i^2 + \sigma^2)}} \exp\left\{-\frac{1}{2}\frac{[y(\nu_i,t_i)
- y_i]^2}{\sigma_i^2 + \sigma^2}\right\},
\end{equation}
where $N$ is the number of measurements, $y(\nu_i,t_i)$ is the above described
integration of Equation 1 against the spectral curve of the $i$th measurement
at the time of the $i$th measurement; $y_i$ is the $i$th measurement in units of log
spectral flux; $\sigma_i$ is the uncertainty in the $i$th measurement in the same
units,
and $\sigma$ is a parameter, sometimes called the slop parameter, that models
the
small systematic errors that are unavoidably introduced when data are
collected from many sources, and other small sources of error (Reichart
2001a).  Ignoring this parameter can lead to erroneous fits and significantly
underestimated uncertainties in the fitted parameter values when the scatter
of the measurements about the fitted model exceeds that which can be accounted
for by the measurement uncertainties alone.

Many of the parameters of the source-frame extinction curve model and all of
the parameters of the Ly$\alpha$ forest absorption and Galactic extinction
curve
models can be constrained a priori.  The source-frame extinction curve model
of Reichart (2001a) is a function of eight parameters:  the source-frame
V-band extinction magnitude $A_V$, $R_V = A_V/E(B-V)$, the intercept $c_1$ and slope $c_2$
of the linear component of the source-frame UV extinction curve, the strength
$c_3$, width $\gamma$ and center $x_0$ of the UV bump component of the
extinction curve,
and the strength $c_4$ of the FUV excess component of the extinction curve.  The
Ly$\alpha$ forest absorption model of Reichart (2001a) is a function of a
single
parameter, $D_A$, the flux deficit.  Reichart (2001a) determines prior
probability distributions for $R_V$, $c_1$, $\gamma$, $x_0$, and $D_A$, which means
that the
values of these parameters can be weighted by fairly narrow distributions, the
description of which sometimes depends on other parameters ($c_2$ and $z$), a
priori.  We adopt these priors here, which can be thought of as increasing the
degrees of freedom by five.  Also, the Galactic extinction curve model of
Cardelli, Clayton \& Mathis (1989) is a function of $E($B$-$V$) = 0.028$ mag for this line of sight (Schlegel, Finkbeiner \& Davis 1998) and a single parameter, $R_V^{MW}$.
We adopt a prior for this parameter that is log normally distributed with mean
$\log 3.1$ and width 0.1, which closely approximates the distribution of values
of this parameter along random lines of sight through the Galaxy (e.g.,
Reichart 2001a; Lee et al. 2002; Galama et al. 2003).   

We fit our model to the data for each of the four standard cases of Sari,
Piran \& Narayan (1998) and Chevalier \& Li (2000), which relate $\alpha_{fs}$ to $\beta_{fs}$ assuming (1)
propagation
through either a constant-density (ISM) or wind-swept (WIND) medium, and (2) a
cooling break that is either redward (RED) or blueward (BLUE) of the observed
optical and NIR frequencies:  For the ISM-RED and WIND-RED cases, $\alpha_{fs} =
(3\beta_{fs}
+ 1)/2 = -(3p - 2)/4$; for the ISM-BLUE case, $\alpha_{fs} = 3\beta_{fs}/2 = -3(p
- 1)/4$; and
for the WIND-BLUE case, $\alpha_{fs} = (3\beta_{fs} - 1)/2 = -(3p - 1)/4$, where $p$
is the
power-law index of the electron-energy distribution.  Since the temporal index
is well constrained by the data, these additional constraints can be powerful
tools for separating the intrinsic spectrum from extinction effects (see
\S4.1).  For purposes of comparison, we also fit our model to the data free of
constraints on $\alpha_{fs}$ and $\beta_{fs}$.

Best fits are found by maximizing the posterior.  Compared to the WIND-BLUE
case, we can rule out the ISM-RED and WIND-RED cases at the 7.3$\sigma$
credible
level, and the ISM-BLUE case is disfavored at the 3.1$\sigma$ credible level.  
Furthermore, the WIND-BLUE fit is consistent with the constraint-free fit,
differing from it at only the 0.6$\sigma$ credible level.  The primary
difference
between these cases is that the WIND-BLUE case requests a shallow intrinsic
spectrum, $\beta_{fs} = -0.34_{-0.01}^{+0.01}$, and a small amount of extinction, $A_V =
0.18_{-0.12}^{+0.25}$ mag, where the other cases request steeper intrinsic spectra
and would fit better if $A_V < 0$ mag were possible (see \S4).  For the WIND-BLUE
case, we find that $\log F_0 = 2.98_{-0.12}^{+0.12}$ $\mu$Jy, $\log t_0 = -2.56_{-0.07}^{+0.07}$
day,
$\alpha_{rs} = -2.16_{-0.10}^{+0.09}$, $\beta_{rs} = 1.1_{-0.8}^{+0.7}$, $\alpha_{fs} =
-1.01_{-0.01}^{+0.02}$, $\beta_{fs} =
-0.34_{-0.01}^{+0.01}$, $A_V = 0.18_{-0.12}^{+0.25}$ mag, $c_2 < 4.3$ (1$\sigma$), and $\sigma =
0.038_{-0.008}^{+0.010}$ mag.\footnote{Due to the dimension of the parameter space, marginalized probability distributions for each parameter value would take impossibly long to compute.  Consequently, these error bars are measured from projected probability distributions and are consequently conservative overestimates.}  The parameters $c_3$ and $c_4$ could not be constrained by
the data.  We plot best-fit light curves for 13 spectral bands in Figure 2 and
best-fit spectral flux distributions for six epochs in Figure 4.  

\subsection{Chromatic Variations}

We plot the residuals of Figure 2 in Figure 3.  One and possibly two significant chromatic variations can be seen from $\approx$40 minutes after the burst until possibly
$\approx$6.0
hours after the burst.  The first of these is an increase relative to the
best-fit model of the unfiltered NEAT and KAIT data, which is also clearly
visible in Figure 2b, concurrent with a decrease relative to the best-fit
model of our R$_{\rm c}$ and possibly I$_{\rm c}$ data from Tenagra and Gunma.  Since the NEAT
and KAIT bandpasses are broad, encompassing the I$_{\rm c}$ and R$_{\rm c}$ bands on their red
ends, this suggests that there was an excess of blue light at this time.  To
explore this further, we plot the spectral flux distribution of the afterglow
in six time slices in Figure 4:

In Figure 4a, we plot the best-fit spectral flux distribution at 67 minutes
after the burst and have scaled all of the data between 39 and 94 minutes
after the burst to this time using the best-fit light curve.  These data
consist of I$_{\rm c}$R$_{\rm c}$VB data from Tenagra and unfiltered NEAT data.  We plot the
weighted average of the scaled NEAT data for clarity.  The combined NEAT point
is only 0.14 mag above the best-fit model, but significantly so, at the
5.2$\sigma$
confidence level.  

In Figure 4b, we plot the best-fit spectral flux distribution at 2.2 hours
after the burst.  We have scaled all of the data between 1.8 and 2.5 hours
after the burst to this time using the best-fit light curve and plot weighted
averages of the scaled data when there are multiple points per spectral band.
 These data consist of a K$_{\rm s}$ point from Fox et al. (2003), I$_{\rm c}$R$_{\rm c}$ data from
Tenagra, an R$_{\rm c}$ point from Gunma, and unfiltered KAIT data.  The combined KAIT
point is even farther above the best-fit model, 0.35 mag, this time at the
3.3$\sigma$ confidence level.  The I$_{\rm c}$ point is now below but still consistent
with
the best-fit model and the R$_{\rm c}$ point is below the best-fit model at the
2.3$\sigma$
confidence level.  Consequently, the KAIT point differs from the R$_{\rm c}$ point at
the 4.0$\sigma$ confidence level with respect to the best-fit model.  Since the
KAIT
bandpass, like the NEAT bandpass, is broad, encompassing the I$_{\rm c}$ and R$_{\rm c}$ bands
on its red end, this suggests that there was an excess of blue light at this
time.  
If we model this excess as an additional power-law component, just in this time slice, we find it to be bluer than $\beta = 1.0$ at the 2$\sigma$ credible level.

In Figure 4c, we plot the best-fit spectral flux distribution at 4.4 hours
after the burst.  We have scaled two points from Fox et al. (2003) -- a B point
at 3.1 hours after the burst and an R$_{\rm c}$ point at 5.7 hours after the burst -- to
this time using the best-fit light curve.  One possibility is that the excess light
has changed from blue to red:  The R$_{\rm c}$ point is above the best-fit model at the
3.1$\sigma$ confidence level and the B point is below the best-fit model at the
3.7$\sigma$
confidence level.  Consequently, these points differ at the 4.8$\sigma$
confidence
level with respect to the best-fit model.  However, given the sparsity of the data in this time slice a temporal variation cannot be ruled out either.

The remaining panels, corresponding to 6.8 -- 11, 17 -- 25, and 46 -- 48 hours
after the burst, show no evidence for significant chromatic variations at
later times. 

Although the third time slice is too sparsely sampled for a temporal variation to be ruled out, the first two time slices, which span the first proposed chromatic variation, are better sampled.  Consider the following simple model:  Let $t_1$ be the beginning of this variation.  Prior to $t_1$, the afterglow is described by Equation 1.  Between $t_1$ and 2.5 hours after the burst, the NEAT and KAIT data are instead described by temporal index $\alpha_{NK}$ and the R$_{\rm c}$ and I$_{\rm c}$ data are instead described by temporal index $\alpha_{R_{\rm c}I_{\rm c}}$.  If this were a temporal variation, $\alpha_{NK}$ would equal $\alpha_{R_{\rm c}I_{\rm c}}$.  Instead, we find that $t_1 = 46_{-21}^{+14}$ min and $\alpha_{NK} - \alpha_{R_{\rm c}I_{\rm c}} = 0.46_{-0.19}^{+0.23}$ with $\alpha_{NK} - \alpha_{R_{\rm c}I_{\rm c}} > 0$ at the 3.5$\sigma$ credible level.  Here, we have fixed all of the other parameters to their previous best-fit values so we can also plot this best fit in Figure 3.  Allowing all of the parameters to vary, we find that $\alpha_{NK} - \alpha_{R_{\rm c}I_{\rm c}} > 0$ at the 3.3$\sigma$ credible level, which again suggests that this is a chromatic variation.

In the above fit, we also find that $\alpha_{R_{\rm c}I_{\rm c}} = -1.30_{-0.26}^{+0.20}$, which is somewhat steeper than the fitted value of $\alpha_{fs}$.  This suggests that the light curve might be steepening during the first and second time slices, and this is consistent with the B point at the beginning of the third time slice also undercutting the model (e.g., Figure 4c).  However, the data are consistent with the model in the fourth, fifth, and sixth time slices, which suggests a minor rebrightening during the third time slice.  Such minor temporal variations are now commonplace -- GRBs 021004 and 030329 are extreme examples -- but further modeling of such variations is beyond the scope of this paper, and frankly beyond the quality of this data set.  However, this does lend some credibility to the possibility that the second variation is temporal instead of chromatic.  A final possibility is that the R point at the end of the third time slice is a statistical variation:  Given 80 points, the probability of encountering a 4.8$\sigma$ variation is 1 in 7900 (ruled out at the 3.8$\sigma$ confidence level).

Finally, we refit the four standard cases to the data, but this time we accommodate the first, chromatic variation with the above simple model and eliminate the second variation, whether chromatic or temporal, by not fitting to the two points of the third time slice.  Compared to the WIND-BLUE case, we now rule out the ISM-RED and WIND-RED cases at the 6.3$\sigma$ credible level and the ISM-BLUE case at the 3.2$\sigma$ credible level.  For the WIND-BLUE case, we find that $\log F_0 = 2.27_{-0.25}^{+0.26}$ $\mu$Jy, $\log t_0 = -2.01_{-0.16}^{+0.17}$ day, $\alpha_{rs} = -1.78_{-0.08}^{+0.07}$, $\beta_{rs} = -0.95_{-0.55}^{+1.45}$, $\alpha_{fs} = -0.89_{-0.04}^{+0.04}$, $\beta_{fs} = -0.26_{-0.03}^{+0.03}$, $A_V = 0.35_{-0.17}^{+0.22}$ mag, $c_2 < 1.6$ (1$\sigma$), $t_1 = 7.5_{-3.5}^{+2.7}$ min, $\alpha_{R_{\rm c}I_{\rm c}} = -0.88_{-0.08}^{+0.08}$, $\alpha_{NK} - \alpha_{R_{\rm c}I_{\rm c}} = 0.12_{-0.06}^{+0.06}$, and $\sigma = 0.028_{-0.007}^{+0.007}$ mag.  The primary difference between this fit and the WIND-BLUE fit of \S3.1 in which these variations are not treated is we now find more source-frame extinction, $A_V = 0.35_{-0.17}^{+0.22}$ mag with $A_V > 0$ mag at the 2.8$\sigma$ credible level.  Also, $\alpha_{R_{\rm c}I_{\rm c}}$ is now consistent with $\alpha_{fs}$, but $\alpha_{NK} - \alpha_{R_{\rm c}I_{\rm c}}$ is still greater than zero at the 3.0$\sigma$ credible level.  

\section{Discussion}

\subsection{Model and Fits}

Our finding that the data are best described by the WIND-BLUE case differs
from the findings of others.  Fox et al. (2003) discount this case in favor of
the ISM-BLUE case, arguing that if the early-time emission is due to a reverse
shock, in a wind-swept medium it is expected to fade quickly and they measure
a slower fading:  $\alpha_{rs} = -1.63 \pm 0.13$.  However, Chevalier, Li \& Fransson
(2004) point out that this measurement depends sensitively on how one
subtracts out (or models) the forward-shock component, arguing that the value
is closer to $\alpha_{rs} = -2.2$.  Using final instead of GCN data, we
find that
$\alpha_{rs} = -2.16_{-0.10}^{+0.09}$ (variations untreated; \S3.1) or $-1.78_{-0.08}^{+0.07}$ (variations treated; \S3.2).  However, in \S4.2 we point out that emission
from
the reverse shock is not necessarily expected to fade quickly in a wind-swept
medium if $A_*$ and other physical parameters are lower than expected, which
appears to be the case for this GRB.

Holland et al. (2004) also adopt the ISM-BLUE case.  The primary difference
between their fit and ours is that we permit source-frame extinction.  When we
fit the ISM-BLUE case, we find that $\beta_{fs} = -0.67$ (variations untreated) or $-0.60$ (variations treated) with $A_V = 0$ mag, which is
very
similar to their fit in a time slice around 0.88 days after the burst: 
$\beta_{fs} =
-0.69 \pm 0.14$ with $A_V$ assumed to be zero.  However, if source-frame extinction
is permitted and one fits to all of the data, we find that the WIND-BLUE case
with a small to moderate amount of source-frame extinction, $A_V = 0.18_{-0.12}^{+0.25}$ mag (variations untreated) or $0.35_{-0.17}^{+0.22}$ mag (variations treated), is
preferred at the 3.1$\sigma$ (variations untreated) or 3.2$\sigma$ (variations treated) credible level.  Figures 4e and 5e can be directly
compared to Figure 3 of Holland et al. (2004).

Finally, Chevalier, Li \& Fransson (2004) adopt the WIND-RED case, guided by
sparse color information that was available at the time, including the two
points of Figure 4c, which we have already identified as discrepant, possibly
due to excess red light at this time (\S3.2).  Permitting source-frame
extinction and fitting to all of the data, we rule this case out at the
7.3$\sigma$ (variations untreated) or 6.3$\sigma$ (variations treated) credible level.

The WIND-BLUE case, however, requires a relatively hard electron energy
distribution -- $p = 1.68_{-0.03}^{+0.01}$ -- so a break at higher energies is
required.
 Bhattacharya (2001) determines the effect of $p < 2$ on the standard equations:
 By introducing a cut-off frequency $\gamma_u$ such that $\gamma_m < \gamma_e < \gamma_u$ and assuming
that $\gamma_u$ evolves directly with the bulk Lorentz factor of the shock, they
find
results similar to the standard prescriptions.  Galama et al. (2003) found a
similar hard electron energy index for GRB 010222, though other ideas, such as
a continuous injection of energy (Bjornsson et al.  2002) or an early
transition to non-relativistic motion (in't Zand et al. 2001; Masetti et al.
2001), have been proposed. 

\subsection{Physical Parameters}

Following the analysis of Chevalier, Li \& Fransson (2004), but for the
WIND-BLUE case, and using the analytic expressions of Granot \& Sari (2002), we
now constrain physical parameters that describe the afterglow and surrounding
medium for a variety of scenarios.  The first constraint comes from the
expression of Granot \& Sari (2002) for the brightness of the afterglow in the
frequency range of our observations, which for the WIND-BLUE case is
max\{$\nu_{sa}$,$\nu_m$\} $< \nu < \nu_c$, where $\nu_{sa}$ is the
self-absorption frequency, $\nu_m$ is the
typical synchrotron frequency, and $\nu_c$ is the electron cooling frequency. This
corresponds to segment G in their Figure 1.  For $p = 1.68$, a luminosity
distance of $d_L = 2.06 \times 10^{28}$ cm (assuming that $\Omega_m = 0.3$,
$\Omega_\Lambda = 0.7$, and $H_0 = 70$
km s$^{-1}$ Mpc$^{-1}$), and an extinction-corrected $F_R$ = 19 $\mu$Jy at 0.1 days after
the
burst, we find:
\begin{equation}
\overline\epsilon_e^{0.68}\epsilon_B^{0.67}A_*E_{52}^{0.67} = 1.02 \times 10^{-5},
\end{equation}
where $\overline\epsilon_e$ is the electron energy fraction when $p < 2$, $\epsilon_B$ is the
magnetic field
energy fraction, $A_*$ measures the density of the wind-swept medium, and $E = E_{52}
\times 10^{52}$ erg is the total energy of the shock, if spherical.  The second
constraint comes from the expression of Granot \& Sari (2002) for $\nu_c(t)$
and the
fact that the data are well described by the WIND-BLUE case even at early
times (see \S4.3).  Taking $\nu_c > \nu_R$ prior to 3.9 minutes after the
burst -- the
time when the forward shock first outshines the reverse shock in the R band -- 
yields:
\begin{equation}
\epsilon_B^{3/2}A_*^2E_{52}^{-1/2} = 6.52 \times 10^{-6}\left(\frac{t_{c,R}}{3.9\,{\rm min}}\right)^{1/2},
\end{equation}
where $t_{c,R}$ is the time that $\nu_c$ passes above the R band.  The third
constraint
is similar to the second in that we take $\nu_m < \nu_R$ prior to 3.9
minutes after
the burst, else the light curve would have faded much more slowly at this
time, as $F_\nu \sim t^{-1/4}$ (Chevalier \& Li 2000; Chevalier, Li \& Fransson
2004):
\begin{equation}
E_{52}^{1/2}\overline\epsilon_e^2\epsilon_B^{1/2} = 1.18 \times 10^{-5}\left(\frac{t_{m,R}}{3.9\,{\rm min}}\right)^{3/2},
\end{equation}
where $t_{m,R}$ is the time that $\nu_m$ passes below the R band.  The final
constraint
comes from the expression of Granot \& Sari (2002) for the brightness of the
afterglow at 8.5 GHz, given that $F_{8.5} < 35$ $\mu$Jy at a mean time of 13
days
after the burst (Fox et al. 2003).  Here we consider four scenarios:  (A)
$\nu_{sa}
< 8.5$ GHz $< \nu_m$, (B) 8.5 GHz $<$ min\{$\nu_{sa}$,$\nu_m$\}, (C)
max\{$\nu_{sa}$,$\nu_m$\} $< 8.5$ GHz, and (D)
$\nu_m < 8.5$ GHz $< \nu_{sa}$.

For scenario A, using the expression of Granot \& Sari (2002) that corresponds
to their segment D, we find:
\begin{equation}
\overline\epsilon_e^{-2/3}\epsilon_B^{1/3}A_*E_{52}^{1/3} = 3.76\times10^{-2} \left(\frac{F_{8.5}}{35\,\mu{\rm Jy}}\right).
\end{equation}
Combining Equations 3, 4, 5, and 6 yields:
\begin{equation}
\overline\epsilon_e = 6.53 \times 10^{-4}E_{52}^{-1}\left(\frac{t_{c,R}}{3.9\,{\rm min}}\right)^{-0.25}\left(\frac{t_{m,R}}{3.9\,{\rm min}}\right)^{0.24},
\end{equation}
\begin{equation}
\epsilon_B = 765 E_{52}^3 \left(\frac{t_{c,R}}{3.9\,{\rm min}}\right)\left(\frac{t_{m,R}}{3.9\,{\rm min}}\right)^{2.04},
\end{equation}
\begin{equation}
A_* = 1.75 \times10^{-5} E_{52}^{-2}\left(\frac{t_{c,R}}{3.9\,{\rm min}}\right)^{-0.5}\left(\frac{t_{m,R}}{3.9\,{\rm min}}\right)^{-1.53},
\end{equation}
\begin{equation}
F_{8.5} = 20 \left(\frac{t_{m,R}}{3.9\,{\rm min}}\right)^{-1.01}\,\mu{\rm Jy}.
\end{equation}
If we additionally require that $\nu_{sa }< 8.5$ GHz $< \nu_m$ at 13 days
after the burst,
using the expressions of Granot \& Sari (2002) for $\nu_{sa}$ and $\nu_m$, we
find that
$F_{8.5} < 6$ $\mu$Jy and $E_{52} > 3.50 \times10^{-3}(F_{8.5}/35\,\mu{\rm Jy})^{1.38}$.  Since the former
of these constraints contradicts Equation 10 for any value of $t_{m,R} < 3.9$ min,
we rule out this scenario.

For scenario B, using the expression of Granot \& Sari (2002) that corresponds
to their segment B, we find:
\begin{equation}
\overline\epsilon_eA_*^{-1}E_{52} = 1.70 \times10^{-4}\left(\frac{F_{8.5}}{35\,\mu{\rm Jy}}\right).
\end{equation}
Combining Equations 3, 4, 5, and 11 yields:
\begin{equation}
\overline\epsilon_e = 0.306 \left(\frac{t_{c,R}}{3.9\,{\rm min}}\right)^{-0.125}\left(\frac{t_{m,R}}{3.9\,{\rm min}}\right)^{1.125}\left(\frac{F_{8.5}}{35\,\mu{\rm Jy}}\right)^{-0.5}, 
\end{equation}
\begin{equation}
\epsilon_B = 7.45 \times10^{-6}\left(\frac{t_{c,R}}{3.9\,{\rm min}}\right)^{0.625}\left(\frac{t_{m,R}}{3.9\,{\rm min}}\right)^{-0.615}\left(\frac{F_{8.5}}{35\,\mu{\rm Jy}}\right)^{1.5}, 
\end{equation}
\begin{equation}
A_* = 3.85 \left(\frac{t_{c,R}}{3.9\,{\rm min}}\right)^{-0.25}\left(\frac{t_{m,R}}{3.9\,{\rm min}}\right)^{0.24}\left(\frac{F_{8.5}}{35\,\mu{\rm Jy}}\right)^{-1}, 
\end{equation}
\begin{equation}
E_{52} = 2.14 \times10^{-3} \left(\frac{t_{c,R}}{3.9\,{\rm min}}\right)^{-0.125}\left(\frac{t_{m,R}}{3.9\,{\rm min}}\right)^{-0.885}\left(\frac{F_{8.5}}{35\,\mu{\rm Jy}}\right)^{0.5}. 
\end{equation}
If we additionally require that 8.5 GHz $<$ min\{$\nu_{sa}$,$\nu_m$\} at 13 days
after the
burst, we find that $t_{m,R} > 2.9$ min, which is technically consistent with $t_{m,R}
< 3.9$ min, and $F_{8.5} > 27$ $\mu$Jy, which is technically consistent with $F_{8.5}
< 35$ $\mu$Jy.  However, this constrains these parameters' values to narrow ranges
and by Equation 15 implies a value for $E_{52}$ that is much too low, given that
the isotropic-equivalent energy in gamma rays alone was $(1.0 \pm 0.1) \times10^{52}$ erg
(Holland et al. 2004) or $1.68_{-0.27}^{+0.32} \times 10^{52}$ erg (Lamb et al. 2004).  Assuming that the efficiency at which energy is
converted to gamma rays is $\sim$20\% (e.g., Beloborodov 2000), then $E_{52} \sim$ many. 
Consequently, we rule out this scenario as well.

For scenario C, using the expression of Granot \& Sari (2002) that corresponds
to their segment G, we find:
\begin{equation}
\overline\epsilon_e^{0.68}\epsilon_B^{0.67}A_*E_{52}^{0.67} = 6.26 \times10^{-5}\left(\frac{F_{8.5}}{35\,\mu{\rm Jy}}\right).
\end{equation}
Combining Equations 3, 4, 5 and 16 yields:
\begin{equation}
\overline\epsilon_e = 6.53 \times10^{-4} E_{52}^{-1}\left(\frac{t_{c,R}}{3.9\,{\rm min}}\right)^{-0.25}\left(\frac{t_{m,R}}{3.9\,{\rm min}}\right)^{0.24},
\end{equation}
\begin{equation}
\epsilon_B = 765 E_{52}^3 \left(\frac{t_{c,R}}{3.9\,{\rm min}}\right)\left(\frac{t_{m,R}}{3.9\,{\rm min}}\right)^{2.04},
\end{equation}
\begin{equation}
A_* = 1.75 \times10^{-5} E_{52}^{-2}\left(\frac{t_{c,R}}{3.9\,{\rm min}}\right)^{-0.5}\left(\frac{t_{m,R}}{3.9\,{\rm min}}\right)^{-1.53},
\end{equation}
\begin{equation}
F_{8.5} = 6\,\mu{\rm Jy}.
\end{equation}
If we additionally require that max\{$\nu_{sa}$,$\nu_m$\} $< 8.5$ GHz at 13 days
after the
burst, we find that $t_{m,R} < 13$ minutes and $E_{52} > 1.27\times10^{-3}$, neither of which
are problematic.  Taking $E_{52}$ to be as low as 4 and $t_{c,R}$ and $t_{m,R}$ to be as low as the
duration of the burst ($T_{90} = 2.41 \pm 0.15$ sec in the 30 -- 85 keV band, in which
$\nu F_{\nu}$ peaks; Crew et al. 2003), yields $\epsilon_B \sim 0.04$.  In this
case, $\epsilon_e \sim 0.0002$ and
$A_* \sim 0.01$.  However, for $\epsilon_B$ to be this low requires considerable fine tuning:  If $E_{52}$ is as high as 11, $t_{c,R}$ is as high as 54 sec, $t_{m,R}$ is as high as 11 sec, or lesser combinations of these three, $\epsilon_B \approx 1$.  Consequently, $\epsilon_B$ is likely considerably more than 0.04, in which case $\epsilon_e$ can be no more than a factor of three greater
and
is likely less and $A_*$ can only be less.

Finally, for scenario D, using the expression of Granot \& Sari (2002) that
corresponds to their segment A, we find:
\begin{equation}
\epsilon_B^{-1/4} A_*^{-1} E_{52}^{3/4} = 8.60 \times10^{-2} \left(\frac{F_{8.5}}{35\,\mu{\rm Jy}}\right).
\end{equation}
Combining Equations 3, 4, 5 and 21 yields:
\begin{equation}
\overline\epsilon_e = 0.232\left(\frac{t_{c,R}}{3.9\,{\rm min}}\right)^{-0.125}\left(\frac{t_{m,R}}{3.9\,{\rm min}}\right)^{0.75}\left(\frac{F_{8.5}}{35\,\mu{\rm Jy}}\right)^{-0.5},
\end{equation}
\begin{equation}
\epsilon_B = 1.71 \times10^{-5}\left(\frac{t_{c,R}}{3.9\,{\rm min}}\right)^{0.625}\left(\frac{t_{m,R}}{3.9\,{\rm min}}\right)^{0.51}\left(\frac{F_{8.5}}{35\,\mu{\rm Jy}}\right)^{1.5},
\end{equation}
\begin{equation}
A_* = 2.21 \left(\frac{t_{c,R}}{3.9\,{\rm min}}\right)^{-0.25}\left(\frac{t_{m,R}}{3.9\,{\rm min}}\right)^{-0.51}\left(\frac{F_{8.5}}{35\,\mu{\rm Jy}}\right)^{-1},
\end{equation}
\begin{equation}
E_{52} = 2.82 \times10^{-3}\left(\frac{t_{c,R}}{3.9\,{\rm min}}\right)^{-0.125}\left(\frac{t_{m,R}}{3.9\,{\rm min}}\right)^{-0.51}\left(\frac{F_{8.5}}{35\,\mu{\rm Jy}}\right)^{0.5},
\end{equation}
If we additionally require that $\nu_m < 8.5$ GHz $< \nu_{sa}$ at 13 days
after the burst,
we find that $t_{m,R} > 2.9$ min, which is technically consistent with $t_{m,R} < 3.9$
min, and $F_{8.5} > 7$ $\mu$Jy, which is consistent with $F_{8.5} < 35$ $\mu$Jy.  Once
again, this constrains these parameters' values to relatively narrow ranges
and by Equation 25 implies a value for $E_{52}$ that is much too low. 
Consequently, we rule out this scenario as well.  

Consequently, we find that $\nu_m < \nu_R < \nu_c$ after $<$3.9 minutes
after the burst and
max\{$\nu_{sa}$,$\nu_{m}$\} $< 8.5$ GHz around 13 days after the burst.  In this
scenario, $\overline\epsilon_e$
and $A_*$ are considerably lower than canonical values.  Since $F_R \propto \overline\epsilon_e^{0.68}\epsilon_B^{0.67}A_*E_{52}^{0.67}$ 
(Equation 3), both of
these
contribute to the faintness of the afterglow (\S1).  

If we allow ourselves to be guided by the standard-energy result (Frail et al.
2001; Bloom, Frail \& Kulkarni 2003), $E_{52}$ is also lower than what one might
expect:  For wind-swept media, the total energy released in gamma rays is
typically measured to be many $\times$ 10$^{50}$ erg (Bloom, Frail \& Kulkarni 2003).  Given
that the isotropic-equivalent energy in gamma rays for GRB 021211 was $(1.0 \pm
0.1) \times 10^{52}$ erg (Holland et al. 2004) or $1.68_{-0.27}^{+0.32} \times 10^{52}$ erg (Lamb et al. 2004), this implies a jet opening/viewing angle
of $\sim$20$^\circ$, which is about three times the canonical value.  Hence, $E_{52}$ is
probably about an order of magnitude less than what one might have expected. 
Consequently, we find that GRB 021211's afterglow is faint for a combination
of 3 -- 4 reasons:  (1) a low fraction of energy in relativistic electrons, (2)
a low density for the wind-swept medium, implying either a low mass-loss rate
and/or a high wind velocity for the progenitor, (3) a wide opening/viewing
angle for the jet, and possibly (4) moderate source-frame extinction (\S 3.2).  

Furthermore, with $\epsilon_B/\overline\epsilon_e > 200$ and  $\epsilon_B$ likely much greater than 0.04 the jet appears to be
significantly far from
equipartition and magnetically dominated.  This is similar to SN 1993J, for
which the magnetic field energy density is $\sim$10$^4$ times the relativistic
particle energy density (Chandra et al. 2004), but dissimilar to SN 1998bw
(Kulkarni et al. 1998) and SN 2002ap (Bjornsson \& Fransson 2004), which appear
to be near equipartition.

These findings are supported by the existence of the bright reverse shock: 
Fox et al. (2003) dismiss the possibility of a wind-swept medium because for
canonical values of $\overline\epsilon_e^{rs}$, $\epsilon_B^{rs}$, $A_*$, and $E_{52}$, $\nu_c^{rs}$ is
expected to be
significantly less than $\nu_m^{rs}$, in which case the reverse shock is
expected to
fade away quickly and not be bright (Chevalier \& Li 2000).  However, this is
not the case when $\overline\epsilon_e^{rs}$, $\epsilon_B^{rs}$, and/or $A_*$ are sufficiently low. 
Taking
$\overline\epsilon_e^{rs} \sim \overline\epsilon_e$ and $\epsilon_B^{rs} \sim \epsilon_B$ and
substituting Equations 17, 18, and 19 into
Equations 45 and 47 of Chevalier \& Li (2000), we find:
\begin{equation}
\nu_c^{rs} / \nu_m^{rs} = 5.56 \times 10^4 E_{52}^3 \left(\frac{t_{m,R}}{3.9\,{\rm min}}\right)^{0.03} \left(\frac{t}{1\,{\rm min}}\right)^2,
\end{equation}
for $X = 0.75$, $\Delta_{10} = 3$, and $\gamma_3 = 0.3$.  For $t = 1$ minute after the
burst,
corresponding to the beginning of the first detection of the reverse shock
(Wozniak et al. 2002), $E_{52}$ need only be greater than $\sim$0.03 for $\nu_c^{rs} \sim \nu_m^{rs}$.

\subsection{Chromatic Variations}

Globally, the data are well described by the model of \S3.1, but superimposed
on this global behavior are small, but significant variations from $\approx$40
minutes
after the burst until possibly $\approx$6.0 hours after the burst (\S3.2).  This model does not explain these variations.  It merely attemps to accommodate them
with a higher value of the slop parameter:  $\sigma = 0.038_{-0.008}^{+0.010}$ mag. 
The
slop parameter is a global measure of the scatter of the data around the
model, beyond what can be accounted for by the data's error bars (\S3.1). 
Furthermore, at least the first of these variations appears to be chromatic, with a relative increase
of blue light with respect to red light around two hours after the burst, and
possibly, but less certainly, a reversal of this two hours later (\S3.2).  
When we modify the model and fit to better treat these variations, $\sigma$ decreases to $0.028_{-0.007}^{+0.007}$ mag (\S3.2).

One potential explanation for these variations is that we have undersampled a light curve that is
varying in time in such a way as to mimic a chromatic effect.  Indeed, the
high signal-to-noise light curves of GRBs 021004 and 030329 revealed temporal
variations and a variety of explanations have been proposed, including
variations in the density of the external medium (e.g., Lazzati et al. 2002),
refreshed shocks (e.g., Granot, Nakar \& Piran 2003), and patchy shells (e.g.,
Kumar \& Piran 2000).  However, none of these occur on a sufficiently short
timescale to explain the variation around two hours after the burst (Figure 3).
However, a temporal variation cannot be ruled out for the possible
reversal of this effect two hours later (\S3.2).

Another potential explanation is that a spectral break, presumably the cooling
break, is passing through our data around 2 -- 4 hours after the burst. 
Although this is difficult to reconcile with the blue excess of Figure 4b,
since the spectrum is supposed to be half of a spectral index steeper blueward
of the cooling break, it is not necessarily inconsistent with the possible red
excess of Figure 4c.  However, the spectra in Figures 4a and 4e would then
differ by half of a spectral index, which would be noticeable.  

Another potential explanation is that we observed a dust echo -- light
scattered by dust into the line of sight and received with a time delay due to
the greater path length.  Waxman \& Draine (2000) and Esin \& Blandford (2000)
originally proposed dust echoes as an alternative explanation for the
supernova-like components to the afterglows of GRB 980326 (Bloom et al. 1999)
and GRB 970228 (Reichart 1999; Galama et al. 2000).  Reichart (2001b) modeled
and computed dust echo light curves and spectral flux distributions and found
that while dust echoes can mimic supernova light curves they cannot mimic
supernova spectral flux distributions, at least not near the spectral peak. 
Moran \& Reichart (2004) take the model of Reichart (2001b) and instead of
applying it to dust shells of inner radius $\sim$10$^{18}$ cm, which is what is required
to mimic supernova light curves, they apply it to dust shells of inner radius
$\sim$10$^{14} - 10^{15}$ cm, which is typical of late-type WC Wolf-Rayet stars, the likely
progenitors of GRBs.  

Moran \& Reichart (2004) find that for (1) wind-swept media, (2) bright optical
flashes -- reverse shocks that outshine the forward shock at early times -- and
(3) wide jet opening angles, dust echoes may be observable on a timescale of
minutes to hours after the burst.  Furthermore, the characteristic signature
will be an excess of blue light (since blue light scatters preferentially)
that quickly transitions to an excess of red light (due to increasing path
lengths through dust with increasing time delay).  Since all of these
conditions appear to be met for GRB 021211 (\S3.1, \S4.1, \S4.2), and excess blue
light is observed on this timescale, as well as a possible transition to
excess red light hours later, we now test this hypothesis in two ways:

\noindent
1.  Equation 6 of Moran \& Reichart (2004) gives the turn-on time of an
idealized, on-axis dust echo as a function of the inner radius $R$ of the
circum-progenitor dust shell and the initial opening angle $\theta_{jet}$ of the
jet. 
Dust echoes should not be visible at very early times because X rays from the
burst and UV light from the optical flash should sublimate the dust within
$\theta_{jet}$ of the jet axis this close to the burst (e.g., Waxman \& Draine
2000;
Fruchter, Krolik \& Rhoads 2001; Reichart 2001c).  Taking the turn-on time to
be $\sim$0.4 of the peak time (Reichart 2001b; Moran \& Reichart 2004) and taking
the peak time for GRB 021211 to be $\sim$2.2 hours, we solve for $R$:
\begin{equation}
R \sim 3 \times 10^{15} \left(\frac{\theta_{jet}}{10^\circ}\right)^{-2}\,{\rm cm}.
\end{equation}
For a wide jet opening angle, this yields $R \sim 10^{14} - 10^{15}$ cm, which is the
expectation if late-type WC stars are indeed the progenitors of GRBs.  However, given that the isotropic-equivalent energy in gamma rays alone was $(1.0 \pm 0.1) \times10^{52}$ erg (Holland et al. 2004) or $1.68_{-0.27}^{+0.32} \times 10^{52}$ erg (Lamb et al. 2004) and that $A_* \la 0.01$ (\S4.2), the deceleration radius $r_d$ was likely greater than $10^{16}$ cm (e.g., Equation 7 of Moran \& Reichart 2005) and consequently a value for $R$ likely cannot be deduced.

\noindent
2.  Equation 5 of Moran \& Reichart (2004), but with $r_d$ substitued for $R$, gives the peak brightness of the
optical flash off of the jet axis, at angles around $\theta_{jet}$, as a function
of the
peak brightness of the dust echo, the deceleration radius, the
optical depth $\tau_{\nu (1+z)}$ through the rest of the dust shell at frequency
$\nu (1+z)$, also at
angles around $\theta_{jet}$, and the duration of the optical flash.  Taking the
peak
brightness of the dust echo to be $\sim$19 mag and the duration of the optical flash to be $\sim$30
seconds (e.g., Sari \& Piran 1999), we find:
\begin{equation}
m_{OF}(\theta \sim \theta_{jet}) \sim 9 +2.5\log \tau_{\nu (1+z)} - 
2.5\log \left(\frac{r_d}{10^{16}\,{\rm cm}}\right)\,{\rm mag},
\end{equation}
Using our best fit (\S3.1), extrapolation of the reverse shock light curve back
to $\approx$30 seconds after the onset of the burst yields R
$\sim 12$
mag.  However, given the known distance to GRB 021211 $m_{OF}(\theta \sim \theta_{jet})$ would have to be significantly fainter lest the dust be sublimated in these directions as well, this close to the burst.  For longer optical flash durations, $\tau_{\nu (1+z)} > 1$, and/or a
greater forward scattering probability than what Esin \& Blandford (2000),
Reichart (2001b), and Moran \& Reichart (2004) assume, $m_{OF}(\theta \sim \theta_{jet})$
would be
fainter, but not sufficiently.  
Alternatively, prior fragmentation of the dust to PAH levels by gamma rays
from the burst might harden it against sublimation, since atomic bonds would then be more difficult to break.

Finally, we point out that neither of these estimates hold in the case of a
jet with a narrow opening angle but a large viewing angle.

\section{Conclusions}

GRB 021211 is one of only a handful of GRBs for which processes other than the
forward shock have been identified at optical wavelengths, which has made it
one of the most studied GRBs.  In this paper, we present additional,
multi-band observations of this event, ranging from minutes to months after
the burst, which in combination with all previously published observations
have allowed us to deeply probe the physics of this GRB and properties of its
circum-progenitor environment.

Coupling the standard afterglow model with a general-purpose extinction curve
model, we find that the afterglow is best described by propagation through a
wind-swept medium, which implies a massive-star progenitor (e.g., Price et al. 2002).  The jet itself appears to be significantly far from equipartition and magnetically dominated.  Indeed, the low fraction of energy
in relativistic electrons appears to be the primary reason that this afterglow
is so faint.  This, combined with a low-density medium, a wide jet
opening/viewing angle, and possibly moderate extinction might be important clues as to why many afterglows are dark/dim in the Swift era, even at early times after the burst.  These findings are supported by the existence of the bright reverse
shock -- in a wind-swept medium this should only be possible if $A_*$ is low
and/or the jet is significantly far from equipartition, meaning that either
$\overline\epsilon_e^{rs}$ or $\epsilon_B^{rs}$ is low as well.

Finally, we observed one and possibly two significant chromatic variations hours after the burst.  We discuss possible reasons for these variations, including the possibility that they are due to a dust echo:  The three primary requirements for an observable dust
echo are a wind-swept medium, a bright optical flash, and a wide jet opening
angle, and the characteristic signature should be an excess of blue light that
quickly reddens, all of which appear to be satisfied for GRB 021211.  However, in the case of GRB 021211 this would imply an off-axis brightness and hence luminosity for the optical flash that would probably sublimate too much dust in these directions.  
Rapid, multi-band, and preferably
simultaneous multi-band observations of future GRBs might shed more light on
this interesting possibility.

\acknowledgements
DER very gratefully acknowledges support from NSF's MRI, CAREER, PREST, and REU programs, NASA's APRA, Swift GI and IDEAS programs, UNC's Junior Faculty Development Award, and especially Leonard Goodman and Henry Cox.  HSP acknowledges support from NASA's SR\&T program and DOE/UC LLNL contract W-7405-Eng-48.  KK acknowledges support from Ministry of Education, Culture, Sports, Science and Technology of Japan grant-in-aid 16740121.  DQL acknowledges support from NASA subcontract NAGW-4690 (HETE-2) and NASA's LTSA program. 


\clearpage

\begin{deluxetable}{lcccc}
\tablecolumns{5}
\tablewidth{0pt}
\tablecaption{FUN GRB Collaboration Observations of the Afterglow of GRB 021211}
\tablehead{\colhead{Date (UTC)} & Mean $\Delta$t & \colhead{Filter} 
& \colhead{Magnitude\tablenotemark{a}} & \colhead{Telescope}}
\startdata
Dec 11.4732\tablenotemark{b} & 2.84 min & R & 15.24 $\pm$ 0.07 & 0.60m Super-LOTIS \\
Dec 11.4751\tablenotemark{b} & 5.63 min & R & 16.26 $\pm$ 0.12 & 0.60m Super-LOTIS \\
Dec 11.4986 & 39.4 min & I$_{\rm c}$ & 18.60 $\pm$ 0.11 & 0.81m Tenagra II \\       
Dec 11.5114 & 57.9 min & R$_{\rm c}$ & 19.52 $\pm$ 0.13 & 0.81m Tenagra II \\      
Dec 11.5239 & 75.9 min & V & 20.06 $\pm$ 0.41 & 0.81m Tenagra II \\
Dec 11.5366 & 94.1 min & B & $>$19.8 & 0.81m Tenagra II \\
Dec 11.5479 & 1.84 hr & I$_{\rm c}$ & 19.99 $\pm$ 0.24 & 0.81m Tenagra II \\
Dec 11.5525 & 1.95 hr & R$_{\rm c}$ & 20.74 $\pm$ 0.42 & 0.81m Tenagra II \\
Dec 11.5566 & 2.05 hr & R$_{\rm c}$ & 20.70 $\pm$ 0.16 & 0.65m Gunma \\
Dec 11.9583 & 11.7 hr & R$_{\rm c}$ & $>$22.0 & 1.34m Tautenburg \\
Dec 11.9744 & 12.1 hr & I$_{\rm c}$ & $>$20.7 & 1.34m Tautenburg \\
Dec 12.3883 & 22.0 hr & i* & 23.02 $\pm$ 0.12 & 3.5m ARC \\
Dec 13.4680 & 47.9 hr & V & 23.0 $\pm$ 0.5 & 1.0m USNO \\
Dec 28.4283 & 17.0 day & i* & 24.41 $\pm$ 0.22 & 3.5m ARC \\
Mar 23.1335 & 102 day & i* & 24.51 $\pm$ 0.29 & 3.5m ARC \\
\enddata
\tablenotetext{a}{Upper limits are 3$\sigma$.}
\tablenotetext{b}{Flux weighted using an iterated power-law index of $\alpha = -1.37$ (\S2.1).}
\end{deluxetable}

\clearpage 

\figcaption[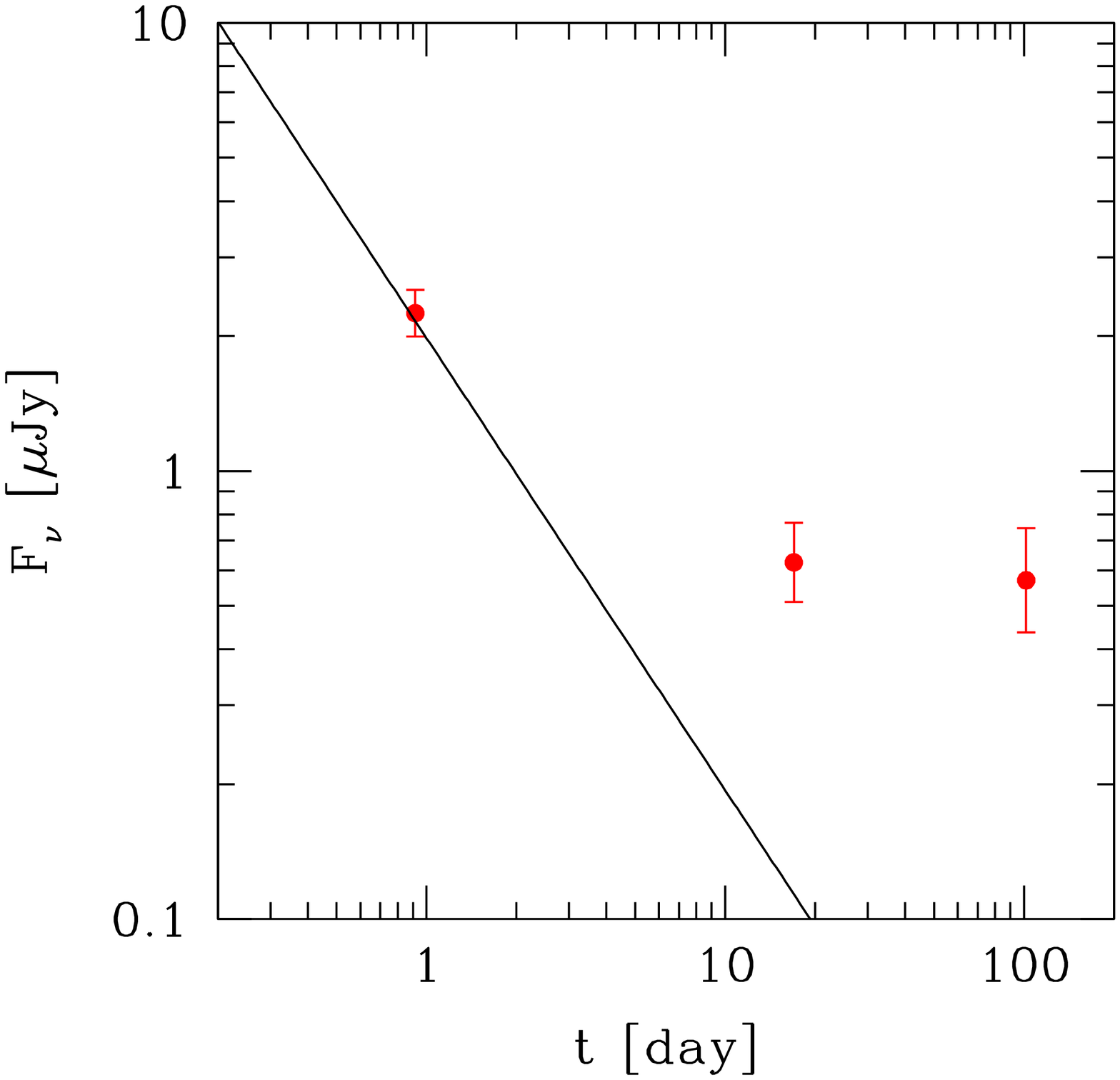]{i$^*$ light curve from 22 hours to 102 days after the burst and
best-fit WIND-BLUE model from \S3.1.  The host galaxy dominates at late times.
 We do not detect the supernova (Fruchter et al. 2002; Della Valle et al.
2003), likely due to the timing of our observations.}

\figcaption[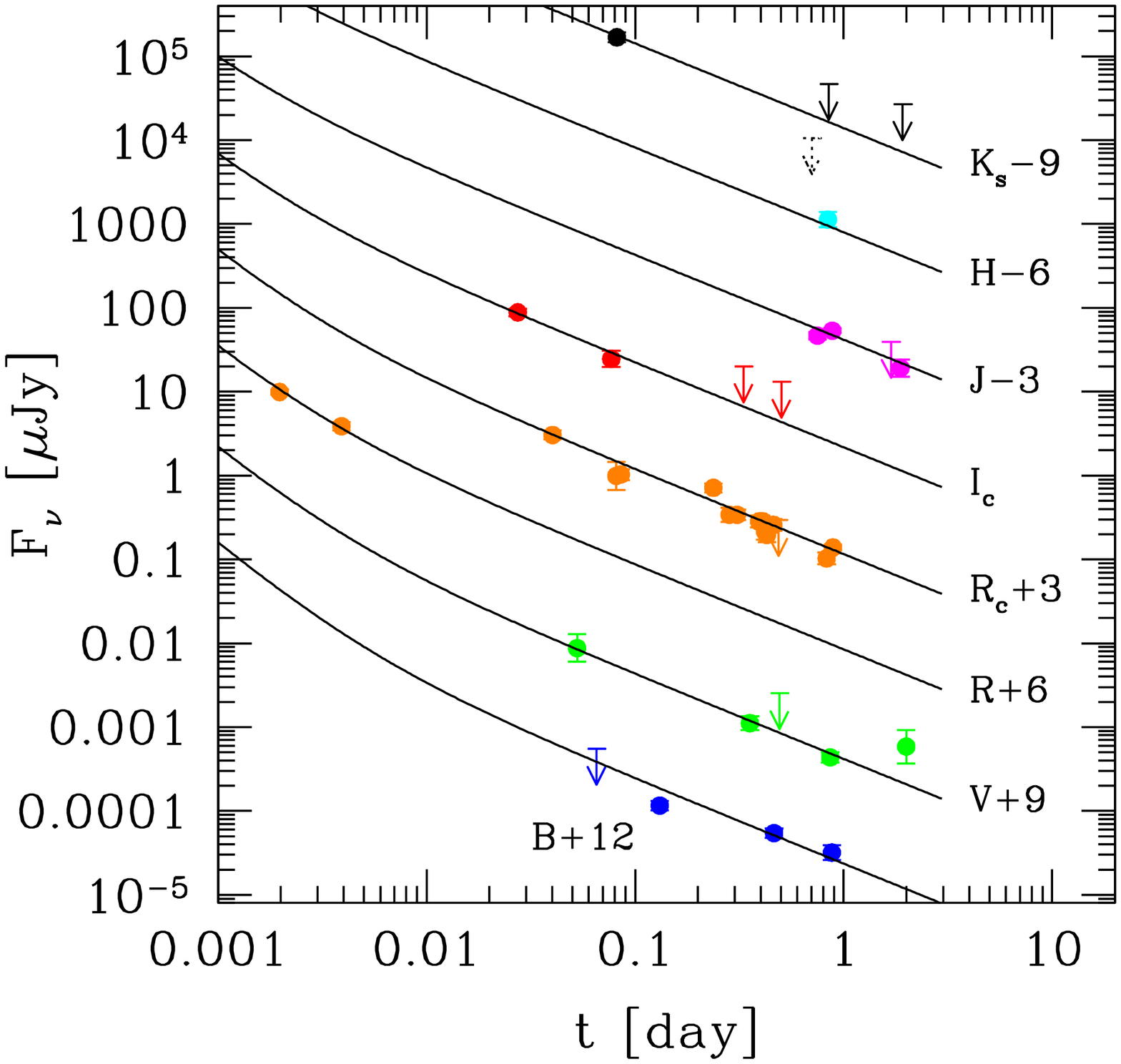]{{\it Top panel:}  BVRR$_{\rm c}$I$_{\rm c}$JHK$_{\rm s}$ light curves from 2.8 min to 2.0 days after
the burst and best-fit WIND-BLUE model from \S3.1.  Upper limits are 3$\sigma$. 
We do
not include the dotted K$_{\rm s}$ upper limit in our fits.  Data are from McLeod et
al. (2002), Pandey et al. (2003), Fox et al. (2003), Holland et al. (2004),
and this paper.  {\it Bottom panel:}  g$^\prime$r$^\prime$i$^*$ and unfiltered KAIT and NEAT light
curves from 9.2 min to 1.0 days after the burst and best-fit WIND-BLUE model
from \S3.1.  The dotted curves are the reverse and forward shock components of the best-fit model for the spectral response of KAIT's CCD.  Data are from Fox et al. (2003), Li et al. (2003), and this paper.}

\figcaption[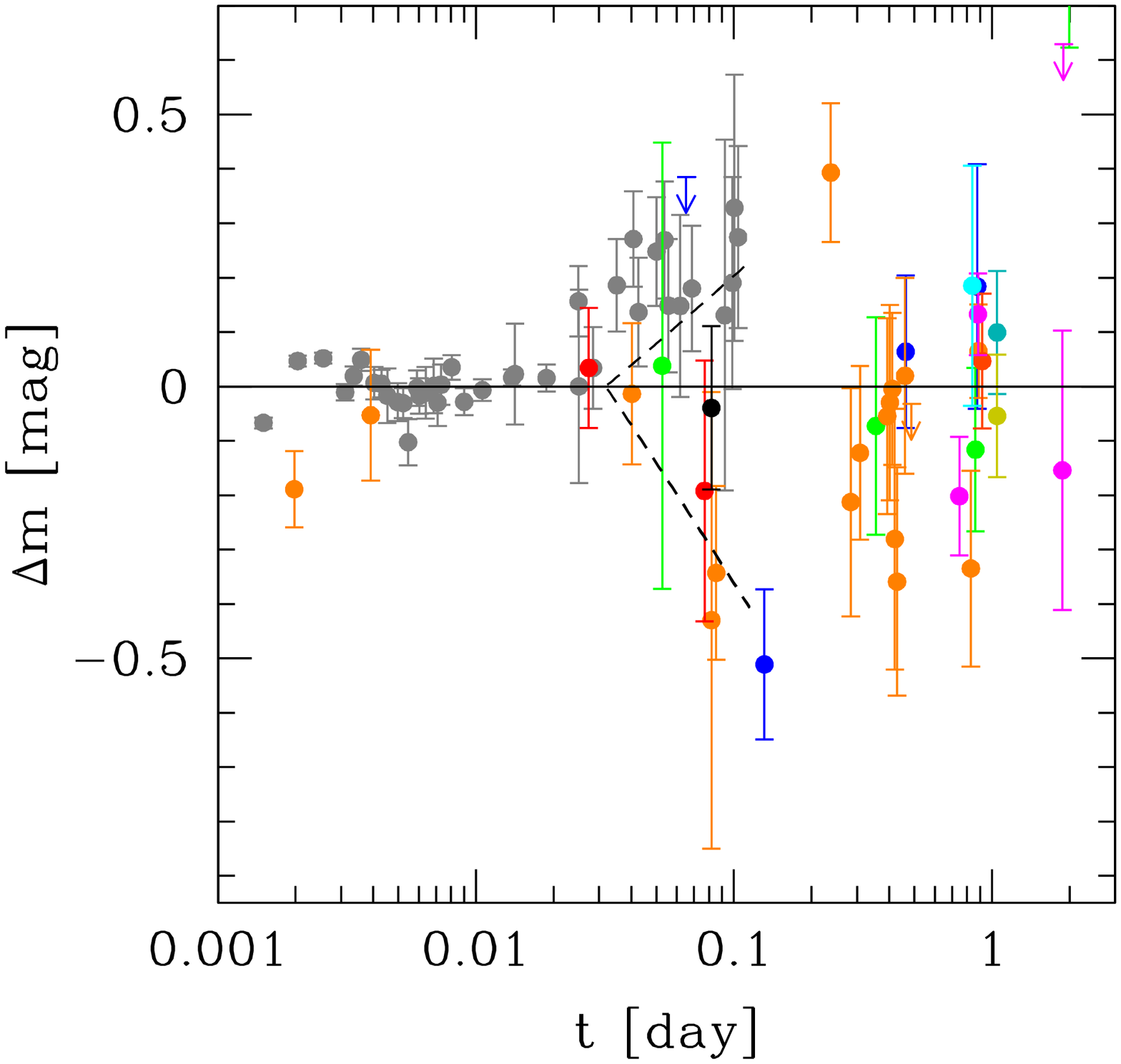]{Residuals of Figure 2.  Colors are the same as in Figure 2.  Notice
the increase relative to the best-fit model of the unfiltered NEAT and KAIT
data, which is also clearly visible in Figure 2b, concurrent with a decrease
relative to the best-fit model of our R$_{\rm c}$ and possibly I$_{\rm c}$ data from Tenagra and
Gunma.  The dashed curves are our best-fit simple model for this from \S3.2.}

\figcaption[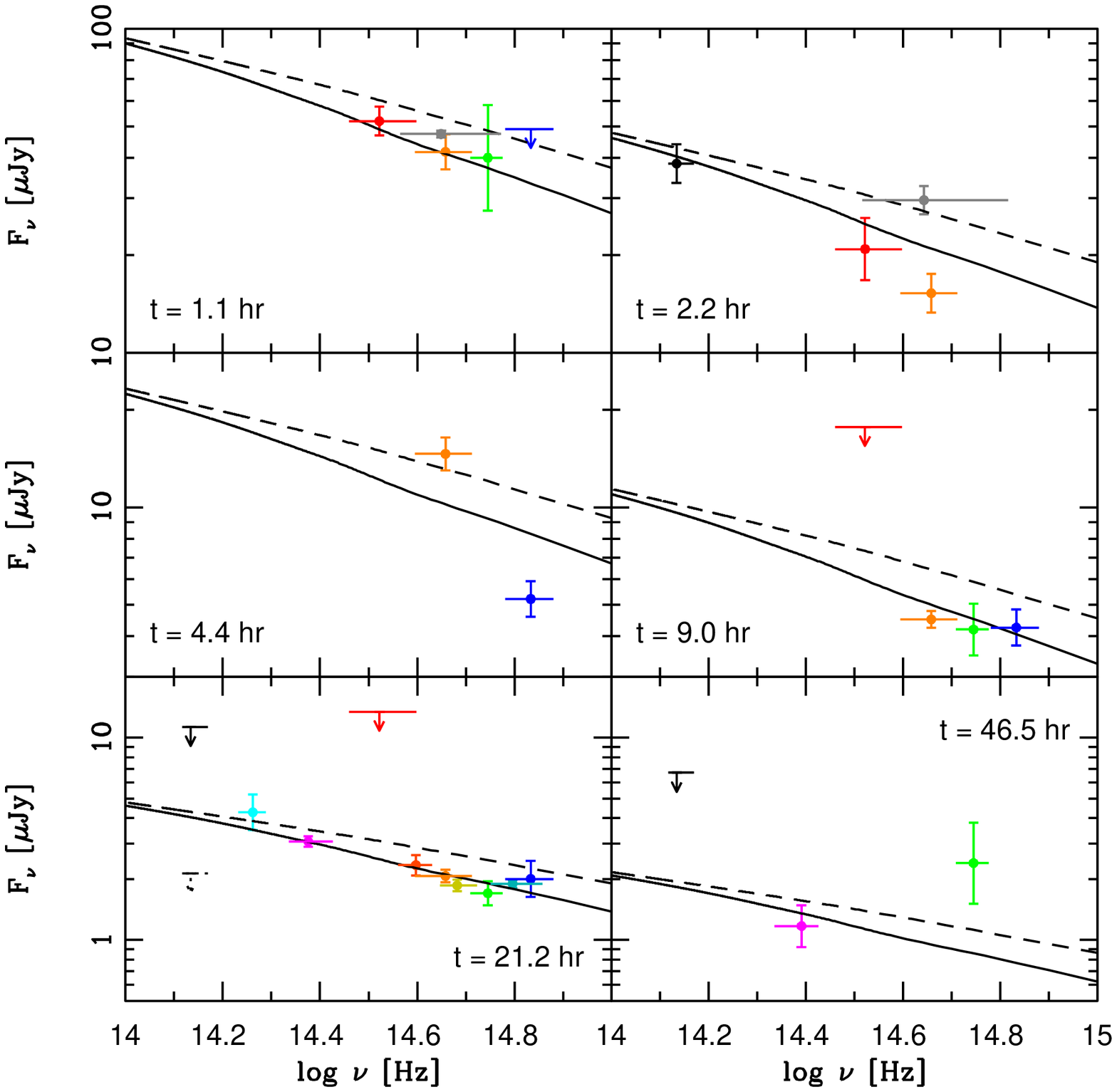]{Spectral flux distribution at six epochs and best-fit WIND-BLUE
model from \S3.1 (solid curves).  Dashed curves are the same fit, but with
source-frame extinction set to zero.  We scale data to these times using the
best-fit light curve and when there are multiple points per spectral band we
plot weighted averages of the scaled data for clarity (see \S3.2).  Colors are
the same as in Figure 2.  Horizontal bars mark the 90\% width of the filter. 
Upper limits are 3$\sigma$.  We do not include the dotted K$_{\rm s}$ upper
limit in
our fits.}

\clearpage

\setcounter{figure}{0}

\begin{figure}[tb]
\plotone{f1.eps}
\end{figure}

\begin{figure}[tb]
\plotone{f2a.eps}
\end{figure}

\begin{figure}[tb]
\plotone{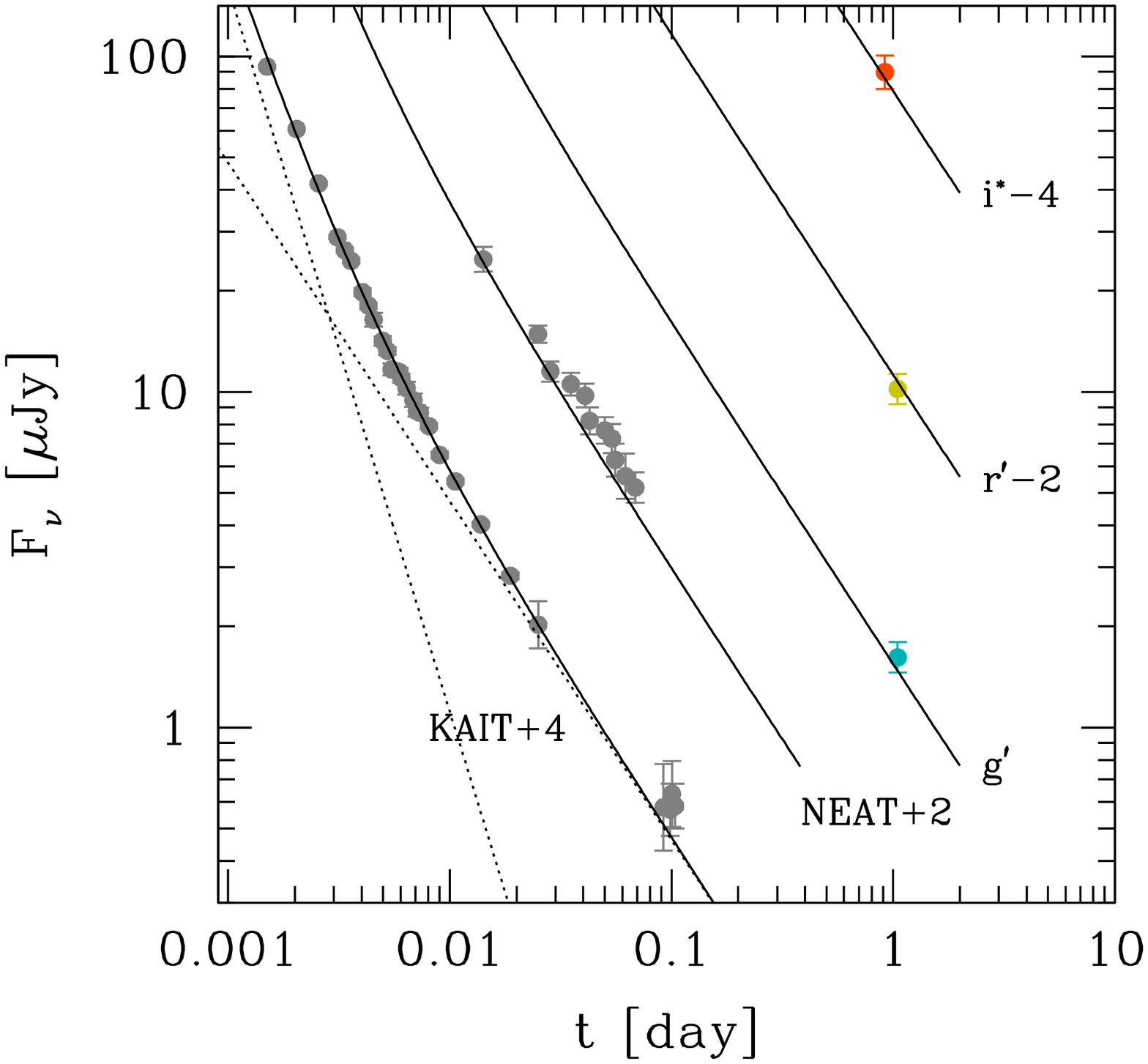}
\end{figure}

\begin{figure}[tb]
\plotone{f3.eps}
\end{figure}

\begin{figure}[tb]
\plotone{f4.eps}
\end{figure}

\end{document}